\documentstyle[prb,aps,epsf,psfrag,graphicx,floats]{revtex}

\newcommand{\ee}{\end{eqnarray}}
\newcommand {\be}[1]{\begin{eqnarray} \mbox{$\label{#1}$}  }

\begin{document}
\twocolumn[\hsize\textwidth\columnwidth\hsize\csname@twocolumnfalse\endcsname

\title{A Carbon Nanotube Based Nanorelay} 
\author{J.~M.~Kinaret, T.~Nord and S.~Viefers}
\address{Department of Applied Physics, Chalmers University of
  Technology \\ and G\"oteborg University, SE-412 96 G\"oteborg,
  Sweden} 
\date{\today} 
\maketitle

\begin{abstract}
  We investigate the operational characteristics of a nanorelay based
  on a conducting carbon nanotube placed on a terrace in a silicon
  substrate. The nanorelay is a three terminal device that acts as a
  switch in the GHz regime. Potential applications include
  logic devices, memory elements, pulse generators, and current or voltage
  amplifiers. \\ \\
\end{abstract}
]
\narrowtext

Nanoelectromechanical systems (NEMS) are a rapidly growing research
field with substantial potential for future applications. The basic
operating principle underlying NEMS is the strong electromechanical
coupling in nanometer-size electronic devices in which the Coulomb
forces associated with device operation are comparable with the
chemical forces that hold the devices together.  Carbon nanotubes
(CNT) \cite{saito1} are ideal candidates for nanoelectromechanical
devices due to their well-characterized chemical and physical
structures, low masses, exceptional directional stiffness, and good
reproducibility.  Nanotube-based NEMS have internal operating
frequencies in the gigahertz range, which makes them attractive for a
number of applications. Recent progress in this direction includes
fabrication of CNT nanotweezers, \cite{kim1,akita1} CNT based random
access memory, \cite{rueckes1} and super-sensitive sensors.
\cite{collins1,adu1}

In this paper we consider another example of CNT based NEMS, a
so-called nanorelay.  This three-terminal device consists of a
conducting CNT placed on a terraced Si substrate and connected to a
fixed source electrode. A gate electrode is positioned underneath the
CNT so that charge can be induced in the CNT by applying a gate
voltage.  The resulting capacitive force between the CNT and the gate
bends the tube and brings the tube end into contact with a drain
electrode on the lower terrace, thereby closing an electric circuit.
We describe the system with a model based on classical elasticity
theory \cite{landau1} and the orthodox theory of Coulomb blockade,
\cite{kulik1,ingold1} and study its IV-characteristics and switching
dynamics. Theoretical studies of a related two-terminal structure have
recently been reported. \cite{dequesnes1,bulashevich1}

\begin{figure}[htbp]
  \begin{center}
    \includegraphics[scale=0.65]{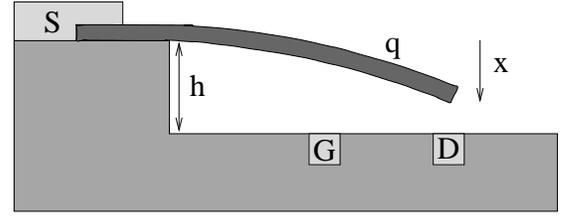} 
    \vspace*{10mm}
    \caption{Schematic picture of the model system consisting
      of a conducting carbon nanotube placed on a terraced Si
      substrate. The terrace height is labeled $h$, and $q$ denotes
      the excess charge on the tube. The CNT is connected to a source
      electrode (S), and the gate (G) and drain (D) electrodes are
      placed on the substrate beneath the CNT at lengths $L$ and $L_G$
      away from the terrace.  The displacement $x$ of the nanotube tip
      is measured towards the substrate.  Typically, $L\approx 50 -
      100$\,nm, $h\approx 5$\,nm.}
    \label{fig:system} 
  \end{center}
\end{figure}

\noindent
{\em Model system.}  The geometry of the nanorelay is depicted in
\mbox{Fig.  {\ref{fig:system}}}. We model the CNT as an elastic
cantilever using continuum elasticity theory: \cite{landau1} Assuming
that only the lowest vibrational eigenmode is excited, and that the
bending profile upon applying an external force is the same as that of
free oscillations, one can express the potential energy of the bent
tube in terms of the deflection $x$ of its tip as $V= kx^2/2$. The
effective spring constant $k$ depends on the geometry of the tube and
is approximately given by $k \approx 3EI/L^3$.  Here $E$ is Young's
modulus, experimentally determined to be approximately $1.2$\,TPa,
\cite{treacy1,wong1} $L$ is the tube length and $I = \pi (D_o^4 -
D_i^4)/64$ its moment of inertia, $D_o$ and $D_i$ being the outer and
inner diameters of the (multi-walled) CNT. The effective mass of the
tube is $m_{eff} = k/\Omega^2 \approx 3M/(1.875)^2 $, where $M$ is the
total tube mass and $\Omega$ its lowest eigenfrequency. \cite{landau1}
It is known experimentally that Q-factors of CNT cantilevers are of
the order of 170-500.\cite{Poncharal1} We model this by a
phenomenological damping force $-\gamma_d \dot{x}$ in the equations of
motion.

For a metallic nanotube, the effective impedance $Z$ is dominated by
the contact resistance at the source contact and is mostly ohmic.  If
a gate voltage is applied, the resulting capacitive force bends the
tube so that a tunneling current can flow to drain, whereas due to the
long tube-gate distance no electrons tunnel between the tube and the
gate electrode. Finite element calculations show that the drain- and
gate capacitances are well approximated by those of parallel plate
capacitors. \cite{johansson1} We thus model them as
\begin{equation}
  \label{eq:caps}
  \begin{array}{rcl}
    C_d(x) &=& \frac{C_0}{1-\frac{x}{h}(1-C_0/C_h)}, \\ \\
    C_g(x) &=& \frac{2C_0}{1-\kappa\frac{x}{h}(1-C_0/C_h)}. 
  \end{array}
\end{equation}
Here $C_0\equiv C_d(x=0)$ is the drain capacitance for a horizontal
tube and $C_h\equiv C_d(x=h)$ is that for the tube in contact with the
drain electrode. The latter can be estimated from experiments.
\cite{tarkiainen1} The constant $\kappa$ $(0 < \kappa < 1)$ accounts
for the fact that the deflection at the position of the gate is
smaller than that of the tip.  The corresponding capacitive forces
take the form $F_c = -(Q^2/2)\, \nabla(1/C(x))$.

The tunneling resistance is modeled as $R_T = R_0
\exp[(h-x)/\lambda]$, where the tunneling length $\lambda$ depends on
the contact material and is typically of the order of $0.5$ \AA, and
$R_0$ is estimated from experimental results. \cite{tarkiainen1}
Following Ingold and Nazarov \cite{ingold1} for the case of zero
temperature and an ohmic environmental impedance $Z$, we determine the
tunneling rate at a given deflection $x$ and source-drain voltage
$V_{sd}$ as
\begin{eqnarray}
  \label{eq:tunrate} 
  \Gamma(x) = \frac{1}{e^2 R_T(x)}\int_0^{eV} dE\, P(E)\, (eV-E) 
\end{eqnarray} 
where $P(E)$, the probability for energy exchange between the
tunneling electron and the environment, is determined
self-consistently from
\begin{eqnarray}
  \label{eq:PE} 
  E P(E) = \frac{2}{g}\int_0^E dE' \, \left[1 + \frac{\pi}{g} 
    \left(\frac{E-E'}{E_c(x)} \right)^2 \right]^{-1} P(E').
\end{eqnarray} 
Here, $E_c(x) = e^2/(2C(x))$ and $g = R_K/Z$ where $R_K = h/e^2
\approx 25.8\, {\rm k}\Omega$ is the von Klitzing constant.

The above model results in a set of coupled non-linear differential
equations describing both the mechanical motion and the current flow
in the system,
\begin{eqnarray}
  \label{eq:eom} 
   m_{eff}\ddot x(t) &=& -kx(t) -\gamma_d \dot{x}(t) \nonumber \\
   & & + F_{c,gate} + F_{c,drain} + F_{contact} \\
   Z\dot q(t) &=& -\frac{q(t)+V_gC_g(x)}{C_g(x) 
     + C_d(x)} + V_s - Z I_t(t) 
\end{eqnarray} 
where $I_t$ is the stochastic tunneling current, and the zero of the
potential has been chosen such that $V_d = 0$.  The contact force
$F_{contact}$ represents the short range forces between the tube tip
and the drain electrode important at large deflections; presently, we
model the surface interactions as elastic collisions.  For a set of
applied voltages, design parameters and initial conditions, we solve
the equations of motion numerically using a fourth order Runge-Kutta
method. The tunneling probability is computed at every time step, and
tunneling events are treated as a (weighted) random process (for
algorithmic details, see Ref.  \onlinecite{isacsson1}).

\begin{figure}[h]
  \begin{center}
    \includegraphics[scale=0.8]{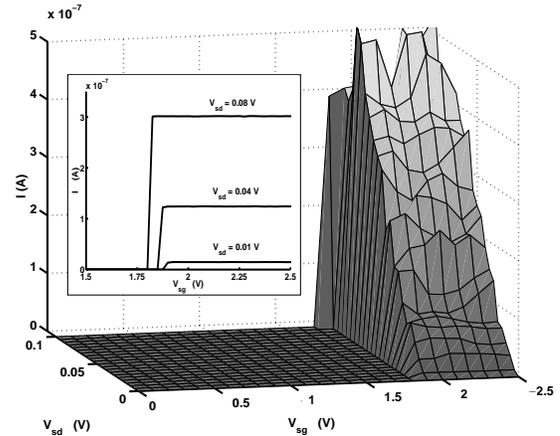}
    \vspace{5 mm}
    \caption{Drain current as a function of source-drain
      voltage and gate voltage. The parameters used here are
      $L=75$\,nm, $h=3$\,nm, $D_i = 2$\,nm, $D_o = 3$\,nm, $L_G =
      0.7\, L$ and $Z = 8 \, {\rm k}\Omega$. The pull-in voltage
      $V_{sg}^*$ for various values of $V_{sd}$ can be read off the
      higher resolution $IV$ curves shown in the inset.}
    \label{fig:iv_vg} 
  \end{center}
\end{figure}

\noindent
{\em Results.}  In our simulations we have studied the dc
IV-characteristics of the nanorelay and its response $I_d(t)$ to a
gate voltage pulse.  \mbox{Fig. \ref{fig:iv_vg}} shows the drain
current as a function of gate- and source-drain voltages.  Due to the
exponential dependence of the tunneling resistance on tube deflection,
there is a sharp transition from a non-conducting (OFF) to a
conducting (ON) state when the gate voltage is varied at fixed
source-drain voltage.  The sharp switching curve allows for
amplification of weak signals superimposed on the gate voltage;
detailed analysis of the amplification characteristics will be pursued
elsewhere.  The inset shows a high resolution plot of the drain
current as a function of $V_{sg}$ for various $V_{sd}$, and
demonstrates that the pull-in voltage $V_{sg}^*$ depends weakly on
$V_{sd}$ as expected, giving rise to highly non-linear $I_d(V_{sd})$
characteristics near the transition.

\begin{figure}[h]
  \begin{center}
    \psfrag{V1}[]{\raisebox{-3mm}{\hspace*{10mm}$V_{sg}$}}
    \psfrag{V2}[]{\raisebox{-1mm}{$V_{sg}$}}
    \psfrag{aa}{$\times 10^{-8}$}
    \psfrag{bb}[]{\raisebox{-0.5mm}{\hspace*{14mm}$\times 10^{-9}$}}
    \psfrag{I (A)}[]{\raisebox{6mm}{\hspace*{-15mm}I (A)}}
    \psfrag{xxx}[]{\hspace{1.5mm}$x$}
    \psfrag{0}{$0$}
    \psfrag{1}{$1$}
    \psfrag{2}{$2$}
    \psfrag{3}{$3$}
    \psfrag{4}{$4$}
    \psfrag{5}{$5$}
    \psfrag{10}{$10$}
    \psfrag{-8}{\small -8}
    \psfrag{-9}{\small -9}
    \includegraphics[scale=0.5]{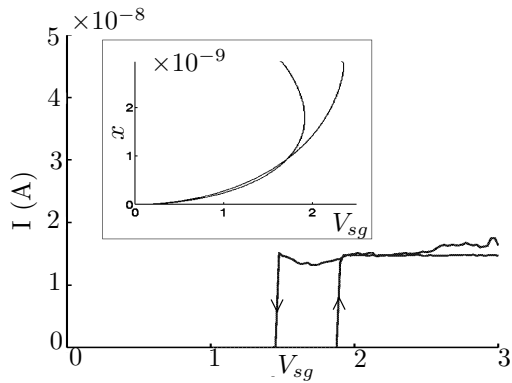}
    \vspace*{5 mm}
    \caption{
      Switching curves for up- and down-sweeps of $V_{sg}$ for various
      $V_{sd}$ showing hysteresis ($L = 75$\,nm, $L_G = 0.7L$, $D_i =
      2$\,nm, $D_o = 3$\,nm).  Inset: contours of zero net force as a
      function of $V_{sg}$ and cantilever deflection $x$ for two
      devices (total force is positive to the right of each curve).
      The appearance of two configurations with zero net force for
      some $V_{sg}$ gives rise to two stable cantilever positions
      corresponding to ON and OFF states, with associated hysteresis.
      Device showing hysteresis $L = 75$\,nm, $L_G = 0.7L$, device
      without hysteresis $L = 100$\,nm, $L_G = 0.5L$, other parameters
      as given above.  }
    \label{fig:hysteresis} 
  \end{center}
\end{figure}

For certain choices of parameters, hysteresis occurs in the OFF-ON-OFF
transition.  This is illustrated in \mbox{Fig. \ref{fig:hysteresis}}
which shows $I_d$ as a function of $V_{sg}$ for up- and down-sweeps.
The hysteresis is due to the appearance of two stable cantilever
positions for certain design parameters and gate voltages: one of the
stable positions is at contact (ON state) and the other is a slightly
bent tube configuration ($x \alt h/2$) corresponding to an OFF state.
The inset of \mbox{Fig. \ref{fig:hysteresis}} shows the contours of
zero net force as a function of gate voltage and tube deflection (net
force is positive to the right of the contour) for two sets of design
parameters; hysteresis arises for the first set of parameters in a
range of gate voltages when there are two zero force configurations.
The amount of hysteresis can be controlled by device design --- for
example, placing the gate electrode further from the drain decreases
hysteresis.  The hysteretic switching characteristics may be utilized
to construct nanotube-based nanoelectromechanical memory elements.

Finally, we studied the switching dynamics of the nanorelay, choosing
a set of parameters for which hysteresis is unimportant.  The
simulation presented here started in the OFF state with $V_{sg}=0$V,
switched to the ON state ($V_{sg}=3$V, see \mbox{Fig.
  \ref{fig:iv_vg}}) at $t = 0.5 \, \mu$s and back to $V_{sg}=0$V at $t
= 1 \, \mu$s.  The resulting switching times were tens of nanoseconds
for the OFF-ON transition ({\em e.g.} rise time $\tau_R \approx
20$\,ns), while the ON-OFF transition was considerably faster (fall
time $\tau_F \approx 0.1$\,ns). This asymmetry is in part due to the
large difference between the step height and the tunneling length, in
part to the cantilever bouncing off the drain surface during the
OFF-ON transition.  The switching dynamics are quite sensitive to the
geometrical parameters which means that they can be optimized for
specific applications. The time constants also depend on the details
of the surface force $F_{contact}$ which are beyond our model.  In
particular, the OFF-ON transition time is greatly reduced by
dissipative surface processes ({\em e.g.} phonon emission) that dampen
tube vibrations, and allow the CNT to settle faster at a stationary
near-contact position.

The van der Waals (vdW) \cite{israelachvili1} and adhesive
\cite{chen1} forces between the CNT and the substrate were neglected
in our model.  Dequesnes {\em et al.} \cite{dequesnes1} recently
studied the effect of vdW forces on the pull-in voltage $V_g^*$ for a
two-terminal device (the entire substrate acting as a gate electrode).
The main result is that vdW forces reduce the pull-in voltage, but do
not change the qualitative features of the OFF-ON transition.
Moreover, the significance of the vdW forces decreases with decreasing
tube length and increasing terrace height or tube diameter, and their
effect on $V_g^*$ is negligible for a wide range of design parameters.
Concerning the ON-OFF transition, adhesive forces \cite{chen1},
enhance the tendency of the CNT to remain in contact even after the
gate voltage has been switched off.  We have performed a simple
analysis comparing the magnitude of the adhesive forces (modeled by a
Morse potential) and vdW forces (based on a Lennard-Jones potential)
to the mechanical and electrostatic ones, and find that the sticking
problem can be avoided by a suitable choice of parameters such as
shorter or thicker CNTs.  Another effect we have neglected is the
possibility of field emission, \cite{bonard1} which may play a role in
the geometry considered in this paper.  An investigation of these
points will be the focus of a future publication.

In conclusion, we have performed a theoretical analysis of a CNT-based
three-terminal nanoelectromechanical switch. The large parameter space
for its design allows for a number of potential applications, such as
amplifiers, logic devices or memory elements.

\bigskip
\noindent
We would like to thank Torgny Johansson, Hans Hansson, Andreas
Isacsson, Herre van der Zant and Eleanor Campbell for useful
discussions.  We gratefully acknowledge financial support from the
Foundation for Strategic Research through its program CARAMEL (Carbon
Allotropes for Microelectronics).

\bibliographystyle{plain}

\end{document}